\documentclass[prb,aps,twocolumn,amsmath,amssymb]{revtex4}

\usepackage{graphicx}
\usepackage{dcolumn}
\usepackage{bm}

\begin{document}
\title{Charge ordering in charge-compensated $Na_{0.41}CoO_2$ by oxonium ions }

\author{C. H. Wang, H. T. Zhang, X. X. Lu and G. Wu }
\author{ X. H. Chen}
\altaffiliation{Corresponding author} \email{chenxh@ustc.edu.cn}
\affiliation{ Hefei National Laboratory for Physical Science at
Microscale and Department of Physics, University of Science and
Technology of China, Hefei, Anhui 230026, People's Republic of
China}
\author{J. Q. Li }
\affiliation{Beijing National Laboratory for Condensed Matter
Physics, Institute of Physics, Chinese Academy of Science, Beijing
100080, People's Republic of China}

\date{\today}

\begin{abstract}

Charge ordering behavior is observed in the crystal prepared
 through the immersion of the $Na_{0.41}CoO_2$ crystal in distilled
 water. Discovery of the charge ordering in the crystal with Na content
 less than 0.5 indicates that the immersion in water brings about the reduction of the
 $Na_{0.41}CoO_2$.  The formal valence of Co changes from +3.59 estimated from the
 Na content to +3.50, the same as that in $Na_{0.5}CoO_2$. The charge
 compensation is confirmed to arise from the intercalation of the
 oxonium ions as occurred in the superconducting sodium cobalt oxide bilayer-hydrate.\cite{takada1}
  The charge ordering is the same as
 that observed in $Na_{0.5}CoO_2$. It suggests that the Co valence
 of +3.50 is necessary for the charge ordering.

\end{abstract}

\maketitle
\newpage

\section{Introduction}

The layered sodium cobaltate $Na_xCoO_2$ has become one of the
focus in research. Na doping leads to change spin 1/2 $Co^{4+}$ to
spinless $Co^{3+}$. Discovery of superconductivity with $T_c\sim$
5 K in $Na_{0.35}CoO_2\cdot1.3H_2O$ \cite{Takada2} makes one
consider that $Na_xCoO_2$ may be a good example like cuprates as a
doped Mott insulator becomes superconducting. Furthermore, the
triangle lattice based cobaltates may also show the possibility to
find some unique novel electronic phases, for example, the
Anderson's triangular lattice RVB phases\cite{Anderson} and the
strong topological frustration.\cite{Baskaran1,Kumar,Qiang-Hua
Wang} $Na_xCoO_2$ system shows many novel properties. Large
thermoelecreic power (TEP) is magnetic field dependent and the TEP
enhancement is believed to be due to spin entropy in
$Na_{0.68}CoO_2$, \cite{wangyayu01} the anomalous Hall signal
shows no saturation up to 500 K,\cite{wangyayu02} the unusual
linear-T resistivity found in low
temperature\cite{wangyayu01,foo,arpes-nacoo} belongs to non-Fermi
liquid behavior. In addition, a so-called "incoherent-coherent"
transition with decreasing temperature is observed in $\rho_c(T)$,
a quasiparticle weight shows up well after a dimensional crossover
in $Na_{0.75}CoO_2$.\cite{arpes-nacoo,Valla}

A characteristic of non-hydrate $Na_xCoO_2$ is the sensitivity of
the electronic states to slight change in x. An insulating state
with anomalous change in thermopower, Hall coefficient and thermal
conductivity occurs at x=0.5.\cite{foo} The crystal structure
involves an ordering of the Na ions into zigzag chains, which
decorate charge ordered chains of Co ions.\cite{huang} Charge
ordering near the commensurate fillings x=1/4 and 1/3 is predicted
by theory.\cite{Baskaran} However, such predicted charge ordering
is not observed so far. Very recently, Takada et al. have
reinvestigated the superconducting sodium cobalt oxide
bilayer-hydrate. It is found that the oxonium ions, $(H_3O)^+$,
can occupy the same crystallographic sites as the Na ions when the
sample of the $Na_{0.4}CoO_2$ is immersed in distilled water and
the Co valence is +3.4.\cite{takada1} Due to the intercalation of
$(H_3O)^+$, the optimal $T_C$ in the Co valence range of 3.24-3.35
is also reported\cite{Milne}. Furthermore, the oxygen
nonstoichiometry is reported for x$\leq$0.7.\cite{Karppinen} The
occupation of oxonium ions in the $Na^+$ layers in superconducting
hydrate $Na_xCoO_2$ and the existence of oxygen vacancies in
$Na_xCoO_2$ make the Co valence to be much lower than the results
that deduced from the Na content. While the Co oxidation state of
about +3.7 is widely used to discuss physical properties, to draw
phase diagram and to make theoretical calculations.
\cite{foo,singh,kumar,honerkamp,Baskaran1} Therefore, it is
important to make out the effect of oxonium ions occupation on the
formal valence of Co and the physical properties. In this paper,
we report that the $Na_{0.41}CoO_2$ crystal with metallic behavior
immersed in distilled water shows an insulating behavior, similar
to that observed in $Na_{0.5}CoO_2$. The charge
 compensation by the intercalation of the
 oxonium ions leads to the change in Co valence from +3.59 to
 +3.50. It suggests that Co valence of +3.50 is necessary for the
 charge ordering, and the $Co^{3+}$ and $Co^{4+}$ ions arrange
 orderly.

\section{Experimental}

 High quality single crystals $Na_{0.7}CoO_2$ were grown using the flux
method. The typical dimensional is about $2 \times 1.5 \times 0.01
mm^3$ with the shortest dimension along the c axis. The
$Na_{0.41}CoO_2$ samples were achieved by Na deintercalation of 5
mg $Na_{0.7}CoO_2$ crystal in 5 ml of 6 M $Br_2$ in acetonitrile
at ambient temperature for one week. The actual Na concentration
was determined by Atomscan Advantage inductively coupled plasma
atomic emission spectrometer (ICP). The content of $OH^-$ was
determined by titration with a standard HCl solution. The
oxidation states of Co in the samples were determined by redox
titration. The $Na_{0.41}CoO_2$ crystal ( 20 mg ) is immersed in
distilled water ( 120 ml ) for 48 hours to yield the hydrated
$Na_xCoO_2$. To study the effect of the water immersion on the
physical properties, the in-plane and out-plane resistivities for
the $Na_{0.41}CoO_2$ crystals were measured by standard four probe
method before their immersion in distilled water. The four
electrodes on the crystal $Na_{0.41}CoO_2$ were made by silver
deposition. Then these crystals with the electrodes were immersed
in distilled water. All crystals were characterized by Rigaku
D/max-A X-Ray diffractometer (XRD) with graphite monochromatized
$CuK_{\alpha}$ radiation ($\lambda$=1.5406 \AA) in the 2$\theta$
range of 10-70$^o$ with the step of 0.02 degree at room
temperature. The crystals were characterized by transmission
electron miscroscopy (TEM). The TEM investigations were performed
on an H-9000NA TEM operating at 300 kV, and a TECNAI 20 operating
at a voltage 200 kV. Thin single crystalline samples for TEM
observations were obtained by peeling them from large single
crystals with a tape, and then mounting them on standard electron
microscopy grids. It is noted that the $Na_xCo_2$ materials
intercalated by water are easily damaged under electron beam due
to the high mobility of Na atoms as well as the weak bonding of
$H_2O$ molecules inside the crystals. We therefore performed all
TEM observations at the low temperature of around 100K. It is
found that radiation damage can be almost eliminated during our
measurements. It should be addressed that all results discussed as
follow are well reproducible.

\section{Results and Discussion}

Figure 1 shows the temperature dependence of resistivity for the
$Na_{0.41}CoO_2$ crystal and $Na_{0.41}CoO_2$ crystal immersed in
water. The resistivity behavior of $Na_{0.41}CoO_2$ is similar to
the previous report\cite{foo}. It is found that the $T^2$ behavior
is observed in in-plane resistivity ( $\rho_{ab}(T)$ ) below about
30 K. A $T^{3/2}$ temperature dependence is observed from 30 to
about 100 K, then a T-linear behavior between about 100 and 220 K.
Above 220 K, $\rho_{ab}(T)$ can be described by $T^n$ with n$<$1
as the case of $Na_{0.7}CoO_2$. $Na_{0.41}CoO_2$ shows a metallic
behavior in the whole temperature range. A striking feature is
that the $Na_{0.41}CoO_2$ crystal immersed in water shows a
divergent behavior in resistivity below about 50 K as shown in
\begin{figure}[h]
\includegraphics[width=8cm]{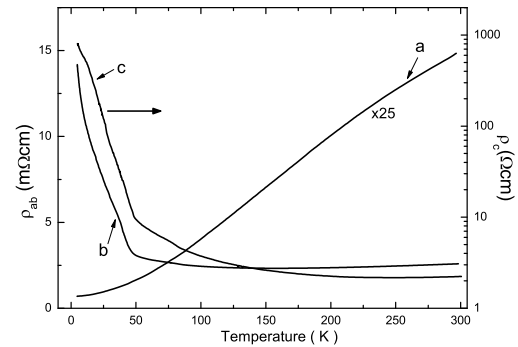}\vspace{-5mm}
\caption{\label{fig:epsart} The temperature dependence of  (a):
in-plane  resistivity for the $Na_{0.41}CoO_2$ crystal; (b):
in-plane  resistivity  and (c): out-of-plane resistivity for the
$Na_{0.41}CoO_2$ crystal immersed in water, respectively.}
\end{figure}
curve b. This behavior is quite similar to that observed in
$Na_{0.5}CoO_2$, in which the resistivity divergency in the low
temperatures is ascribed to the charge ordering.\cite{foo} It
suggests that the charge ordering occurs in the $Na_{0.41}CoO_2$
crystal immersed in water. In order to confirm that the charge
ordering behavior is intrinsic, the out-of-plane resistivity
$\rho_c(T)$ is also measured and the result is shown in curve c.
The out-of-plane resistivity shows a similar divergent behavior to
that of in-plane resistivity, and the temperature corresponding to
the divergency is the same as that observed in $\rho_{ab}(T)$.
These results indicate that the charge ordering is bulk behavior
in $Na_{0.41}CoO_2$ crystal immersed in water. Therefore, the
natural question is why the charge ordering occurs in the
$Na_{0.41}CoO_2$ crystal immersed in water and what happens when
the $Na_{0.41}CoO_2$ is immersed in the distilled water. However,
the charge ordering behavior can be only observed at x=0.5 in the
phase diagram of $Na_xCoO_2$ system.\cite{foo}

In order to compare the charge ordering observed in the
$Na_{0.41}CoO_2$ crystal immersed in water with that appeared in
$Na_{0.5}CoO_2$, the derivative of in-plane resistivity for the
$Na_{0.41}CoO_2$ crystal immersed in water as a function of the
temperature is shown in Fig.2. A dip is observed at about 85 K,
the derivative rapidly decreases at 51 K and 32 K, respectively.
These behaviors are almost the same as that observed in
$Na_{0.5}CoO_2$, in which the features coincide with the cusps
occurred in susceptibility.\cite{huang} It suggests that the
\begin{figure}[h]
\includegraphics[width=8cm]{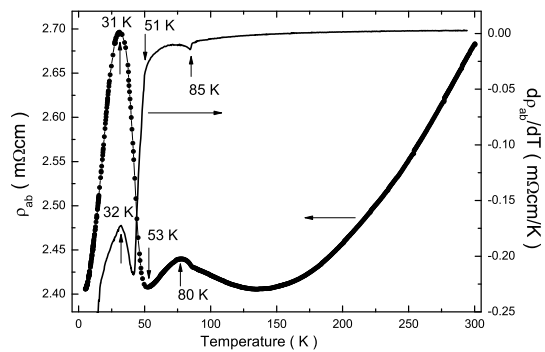}\vspace{-5mm}
\caption{\label{fig:epsart} The derivative of in-plane resistivity
as a function of temperature for the $Na_{0.41}CoO_2$ crystal
immersed in water; The temperature dependence of the in-plane
resistivity for the same crystal after baking at 100 $^0C$.}
\end{figure}
$Na_{0.41}CoO_2$ crystal immersed in water shows the same charge
behavior as in $Na_{0.5}CoO_2$. To make clear the effect of the
immersion on the transport properties, the $Na_{0.41}CoO_2$
crystal immersed in water was baked at 100 $^oC$ for 2 days. The
temperature dependence of in-plane resistivity for the baked
crystal is shown in Fig.2. It shows a metallic behavior above
about 130 K, then slight insulating behavior. A peak shows up
around 80 K, and a weak diverging behavior is observed at 53 K. A
intriguing feature is that an insulator-metal transition occurs at
about 31 K. This is in contrast to the behavior observed in the
$Na_{0.41}CoO_2$ crystal immersed in water or $Na_{0.5}CoO_2$, in
which another rapid divergent behavior is observed at about 30 K
as the behavior of $d\rho_{ab}/dT$ shown in Fig.2. It suggests
that some change has taken place for the baking. In one word, a
charge ordering behavior is found in $Na_xCoO_2$ with x $<$ 0.5,
and is the same as that observed in $Na_{0.5}CoO_2$.

To understand what happened in the $Na_{0.41}CoO_2$ crystal
immersed in the distilled water, all crystals discussed above are
characterized by X-ray diffraction measurement. The x-ray
diffraction patterns for the crystals are shown in Fig.3. Figure 3
shows that only (00l) diffraction peaks are observed. No impurity
peak is observed and the structure remains unchanged except for
the slight change in the c-axis lattice parameter for the crystal
immersed in distilled water. It suggests that the superconducting
hydrated
\begin{figure}[h]
\includegraphics[width=9cm]{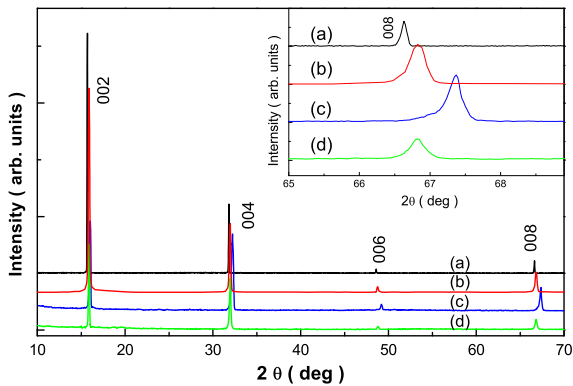}\vspace{-5mm}
\caption{\label{fig:epsart} The X-ray diffraction patterns for
(a): the $Na_{0.41}CoO_2$ crystal; (b): the $Na_{0.41}CoO_2$
crystal immersed in water; (c): the baked crystal (b); (d): the
crystal (c) re-immersed in distilled water. Inset: enlarge the
$2\theta$ range of 65 to 69 degree to clearly show the change in
c-axis lattice parameter.}
\end{figure}
$Na_{0.35}CoO_2\cdot1.3H_2O$ is not formed.\cite{Takada2} This
could be because the time for the immersion of the crystal in
distilled water is not enough for the intercalation of the water
to form the superconducting phase. It is confirmed by the
observation of superconductivity for the immersion of the crystal
in distilled water for more than two months.  For the
$Na_{0.41}CoO_2$ crystal, the c-axis lattice parameter is 11.218
\AA. The lattice parameter of the $Na_{0.41}CoO_2$ crystal
immersed in the distilled water ( 11.203 \AA ) is slightly less
than that of $Na_{0.41}CoO_2$ crystal. While a remarkable change
in the lattice parameter ( about 0.1 \AA ) is observed by baking
at 100 $^oC$ for 48 hours. It is considered that some crystal
water is gone for the baking, this assumption is confirmed by the
re-immersion of the baked crystal in distilled water for 48 hours
as shown in the curve (d) of Fig.3. The c-axis lattice parameter
for the crystals is listed in the table 1. The $Na_{0.41}CoO_2$
crystals immersed in distilled water and baked at 100 $^oC$ for 48
hours are denoted as immersed and baked crystals in table 1,
respectively.

\begin{table*}[htp]
\caption{\label{tab:table1}c-axis lattice parameters, chemical
compositions and oxidation state of Co for the different crystals
}
\begin{ruledtabular}
\begin{tabular}[b]{cccccc}
  & c-axis lattice &  &   &  & oxidation \\
crystal & parameter (\AA) &  Na (wt\%) & Co (wt \%)& x & state of Co\\
\hline\
$Na_{0.41}CoO_2$  & 11.218(3) & 9.30(3) & 57.8(1) & 0.412(2) & 3.60(2) \\\
immersed crystal  & 11.203(3) & 7.21(3) & 53.85(1) & 0.343(3) & 3.49(1)
\\\
baked crystal & 11.114(3) &\\

\end{tabular}
\vspace*{-2mm}
\end{ruledtabular}
\end{table*}

In $Na_{0.5}CoO_2$ with charge ordering, the crystal structure
involves an ordering of the Na ions into zigzag chains, and the
chains are believed to decorate charge-ordered chains of the Co
ions, giving a basis for the highly unusual properties
observed.\cite{huang} In order to confirm it, the $Na_{0.41}CoO_2$
crystal and the $Na_{0.41}CoO_2$ crystal immersed in the distilled
water with charge ordering  are characterized by electron
diffraction. The results are shown in Fig.4. Figure 4(a) shows
[001] diffraction pattern for the $Na_{0.41}CoO_2$ crystal. The
compound shows a superstructure. The superstructure reflections
can be described by an incommensurate vector {\bf q}, oriented
along [110]. Which is the same as that reported by Zandbergen et
al. in $Na_{0.35}CoO_2$.\cite{zandbergen} The $Na_{0.41}CoO_2$
crystal immersed in the distilled water shows different
superstructure, and the typical [001] diffraction pattern is shown
in Fig.4(b). It should be pointed out that the patterns shown in
Fig.4 are taken with only a
\begin{figure}[h]
\includegraphics[width=8cm]{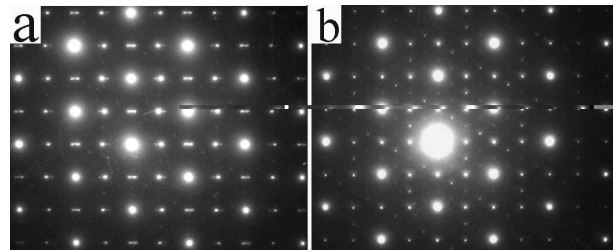}
\caption{\label{fig:epsart} [001] diffraction patterns of (a): the
$Na_{0.41}CoO_2$ crystal and (b) the $Na_{0.41}CoO_2$ crystal
immersed in water. The patterns were taken almost immediately.}
\end{figure}
relatively low exposure to the electron beam. The superstructure
is the same as that observed in $Na_{0.5}CoO_2$,\cite{zandbergen}
and the superstructure reflections can be described by a
commensurate vector {\bf q}, oriented [110] and having a length of
0.25 [110]. It further indicates that the charge ordering in the
$Na_{0.41}CoO_2$ crystal immersed in the distilled water is the
same as that in $Na_{0.5}CoO_2$. So far, it has been found that
only charge ordering occurs in $Na_{0.5}CoO_2$, in which the
insulating state is believed to arise from the interaction between
the charge carriers and the Na ions, and at Na content of 0.5 the
density of states of the carriers is sharply peaked.\cite{foo}
Therefore, the valence of Co ions with +3.50 is necessary to
observe the charge ordering.

The question is how to compensate the charge to make the valence
of Co ions from +3.59 to +3.50 when the $Na_{0.41}CoO_2$ crystal
is immersed in distilled water. Very recently, Takada et al.
reported that the oxonium ions occupy the same crystallographic
sites as the $Na^+$ ions and the composition of the
superconducting phase was determined to be
$Na_{0.343}(H_3O)_{0.237}CoO_2 \bullet 1.19H_2$.\cite{takada1}
When the $Na_xCoO_2$ crystal is immersed in the distilled water,
the two reactions could take place as reported by Takada et
al.\cite{takada1}
\begin{multline}
Na_{0.41}CoO_2 +0.14H_2O \longrightarrow \\
Na_{0.34}(H_3O)_{0.07}CoO_2+0.07Na^+ +0.07OH^-
\end{multline}
\begin{multline}
Na_{0.34}(H_3O)_{0.07}CoO_2+0.12H_2O\longrightarrow\\
Na_{0.34}(H_3O)_{0.15}CoO_2+0.02O_2
\end{multline}

 In reaction (1), the decrease
in the Na content could be explained by partial ion-exchange of Na
ions in the cobalt oxide with protons or the oxonium ions. The
decrease content of Na ion is determined by ICP, that is: the
decease content of Na ions is determined by detecting Na ions in
the aqueous filtrate in which the $Na_{0.41}CoO_2$ crystal has
been immersed. In addition, the Na content for the
$Na_{0.41}CoO_2$ crystal immersed in distilled water for 120 hours
was determined by ICP. It is found that the ratio of Na to Co is
0.343(3):1, consistent with the Na loss by determining Na ions in
the aqueous filtrate. It indicates that the Na content  of the
crystal with charge ordering is 0.34 per formula. It should be
pointed out that the content of Na ion for ion-exchange is the
same as that reported by Takada et al.\cite{takada1} It suggests
that the content of Na ions for ion-exchange is limited. There
exists a discrepancy for the $Na_{0.41}CoO_2$ crystal immersed in
water between the oxidation state of Co calculated from the $Na^+$
and $H_3O^+$ content (+3.59) and that (+3.50) expected for the
appearance of charge ordering. This suggests that the immersion in
water brought about the reduction of the host. As discussed by
Takada et al.,\cite{takada1} the reduction of the host occurs by
accompanying with the intercalation of $H_3O^+$ ions instead of
$Na^+$ ions. This is because the intercalation of $Na^+$ will
consume the $Na^+$ ions in the solution and decrease its pH, while
the lower the pH, the higher the potential of oxygen
evolution,\cite{li} which suppresses the reduction of $Na_xCoO_2$.
As pointed out by Takada et al.,\cite{takada1} it is not clear
that the reaction (1) and (2) occurred sequentially or
simultaneously. Reaction (1) leads to the substitution of the
$H_3O^+$ ions for $Na^+$, while Reaction (2) involves reductive
incorporation of the additional $H_3O^+$ ions.

 In order to confirm the occurrence of the host reduction (or Reaction (2)),
  the oxidation states of Co ions in the
$Na_{0.41}CoO_2$ crystal immersed in water were determined by
redox titration as described in Ref. 1. To get precise oxidation
state and avoid the moisture to affect the precision, the mass of
the crystal (18.126 mg) was weighed by high precise  (1 $\mu g$)
balance in glove box, in which the $H_2O$ content is less than 1
ppm, after the crystal was vacuumed for 12 hours to remove the
moisture from the crystal. It should be mentioned that the
resistivity behavior is the same before the vacuum and after. It
suggests that the water in the structure is stable under vacuum at
room temperature. The result obtained by redox titration indicates
that the oxidation state of Co is +3.49(1). It indicates that the
Reaction (2) takes place when the  $Na_{0.41}CoO_2$ crystal was
immersed in water and the host was reduced. The chemical
composition data and the oxidation state of Co are listed in the
table 1, too. It is found that the Na content, x, decreases from
0.412(2) to 0.343(3) after the immersion. The oxidation state of
Co ion should be (4.0-x) according to chemical formula
$Na_xCoO_2$. The $Na_{0.41}CoO_2$ crystal follows this relation as
shown in table 1. While the oxidation state determined by redox
titration for the $Na_{0.41}CoO_2$ crystal immersed in water for
120 hours is 3.49. It should be pointed out that the results
listed in table 1 is well reproducible. It supports that the the
oxidation state of Co with +3.50 is necessary to observe the
charge ordering. The $H_3O^+$ ions is monovalent, the same as the
$Na^+$ ion. Therefore, it can be reasonably expected that the
$H_3O^+$ ions are accommodated in the $Na^+$ sites. The same
superstructure in the $Na_{0.41}CoO_2$ crystal immersed in water
as in $Na_{0.5}CoO_2$ crystal has given a definite evidence for
the accommodation of $H_3O^+$ ions in  the $Na^+$ sites. In
addition, the c-axis lattice parameter of
$Na_{0.34}(H_3O)_{0.15}CoO_2$ crystal is 11.203 \AA, which is
larger than that (11.125 \AA) of $Na_{0.5}CoO_2$. This discrepancy
should arise from the substitution of large $H_3O^+$ ions for
small $Na^+$ ions. After $Na_{0.34}(H_3O)_{0.15}CoO_2$ was baked
at 100 $^oC$ for 48 hours, its lattice parameter changes to 11.114
\AA, slightly less than that (11.125 \AA) of $Na_{0.5}CoO_2$. We
assume that the baking results in the decomposition of $H_3O^+$
ions into $H^+$ and $H_2O$. This assumption is confirmed by the
mass loss after the $Na_{0.34}(H_3O)_{0.15}CoO_2$ was baked. The
mass loss is evaluated by weighing their mass with high precise (
1 $\mu g$) balance before and after the baking. After the
$Na_{0.41}CoO_2$ crystal immersed in water was vacuumed for 12
hours to remove the moisture in glove box, the mass was weighed to
be 2.221 mg. After baking at 100 $^oC$ for 48 hours in glove box,
the mass of the crystal decreases to 2.162 mg. The mass loss
corresponds to the mass of 0.15 $H_2O$ per the molecular formula
$Na_{0.34}(H_3O)_{0.15}CoO_2$. This result coincides with the
intercalation of 0.15 $H_3O^-$ per formula determined by redox
titration. It should be pointed out that the resistivity behavior
does not change after the vacuum. However, the charge ordering
behavior is destroyed by the baking at 100 $^oC$. It suggests that
the baking results in the removal of the water in structure, while
the vacuum has no effect on the water in structure. Therefore, it
is easily understood that the c-axis lattice parameter (11.114
\AA) of the baked $Na_{0.34}(H_3O)_{0.15}CoO_2$ is less than
11.125 $\AA$ of the $Na_{0.5}CoO_2$ because the baked
$Na_{0.34}(H_3O)_{0.15}CoO_2$ should be $Na_{0.34}H_{0.15}CoO_2$,
that is:  the large  $Na^+$ ions was partially substituted by the
small $H^+$ relative to the $Na_{0.5}CoO_2$. Another evidence for
it is that the c-axis lattice parameter returns to 11.203 $\AA$
after the baked crystal is re-immersed in distilled water for 48
hours as shown in Fig.3. So far, it is well understood that the
$Na_xCoO_2$ crystal with x$<$0.5 shows the insulating behavior in
the low temperature after the $Na_{0.41}CoO_2$ crystal is immersed
in water.
\begin{figure}[h]
\includegraphics[width=8cm]{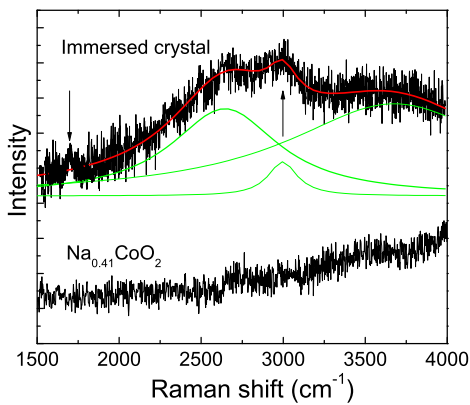}
\caption{\label{fig:epsart} Raman spectra for $Na_{0.41}CoO_2$ and
 $Na_{0.34}(H_3O)_{0.15}CoO_2$ in the wave number range from 1500 $cm^{-1}$ to 4000 $cm^{-1}$.
 The solid lines is the fitting results. }
\end{figure}

In order to confirm the existence of the $H_3O^+$ ion,  Raman
meaurement was carried out with a LABRAM-HR Confocal Laser
MicroRaman Spectrometer using 514.5 nm line from an argon-ion
laser. Figure 5 shows the Raman spectra for the $Na_{0.41}CoO_2$
and  $Na_{0.34}(H_3O)_{0.15}CoO_2$ in the wave number range from
1500 $cm^{-1}$ to 4000 $cm^{-1}$. The
$Na_{0.34}(H_3O)_{0.15}CoO_2$ shows a band at 1700 $cm^{-1}$ and
broad peaks overlapping each other in the wave number range from
2300 $^{-1}$ to 4000 $cm^{-1}$, which is similar to that observed
in the superconducting sodium cobalt oxide
bilayer-hydrate.\cite{takada1} For comparison, the Raman spectra
for $Na_{0.41}CoO_2$ is also shown in Fig.5. No feature is
observed in the wave number range above 1500 $cm^{-1}$. The band
at 1700 $cm^{-1}$ has been ascribed to a bending mode of the
$H_3O^+$ ion,\cite{nash} which can be distinguished from that of
the $H_2O$ molecule whose bending mode at about 1600
$cm^{-1}$.\cite{wilkins} The broad peaks in the wave number range
from 2300 $^{-1}$ to 4000 $cm^{-1}$ were decomposed into three
modes. The most distinct peak at 2995 $cm^{-1}$  can be assigned
to the stretching mode of the $H_3O^+$.\cite{wilkins}. These
results indicate that the crystal immersed in water for 120 hours
definitely contains a considerable amount of the the $H_3O^+$ ion.

\section{Conclusions}

We report that a charge ordering occurs in $Na_{0.41}CoO_2$
compound with x less than 0.5 for first time. The charge ordering
is the same as that observed in $Na_{0.5}CoO_2$. The charge is
compensated by the oxonium ($H_3O^+$) ions, which can be
intercalated into $Na_xCoO_2$ crystal by simple immersion of the
$Na_xCoO_2$ crystal in distilled water and occupy the $Na^+$
crystallographic sites. The oxidation state of Co is +3.50, the
same as that in $Na_{0.5}CoO_2$. It suggests that the
 valence of Co ions with +3.50 is critical to observe the charge
 ordering. In addition, it provides another way to dope the charge carrier
 in conducting $CoO_2$ layers.  It further supports the intercalation of the oxonium
 ions into the superconducting phase: the hydrated $Na_{0.35}CoO_2$.
 Which will lead to different carrier density of the $CoO_2$ layers in
 superconducting phase from 3.7 estimated from the Na content. It
 will challenge the phase diagram for superconducting phase and
 its theory.

\vspace*{-2mm}
\section{Acknowledgment}
We would like thank N. L. Wang and J. L. Luo for helpful
discussions. This work is supported by the Nature Science
Foundation of China and by the Knowledge Innovation Project of
Chinese Academy of Sciences.

\end{document}